\documentstyle[sprocl]{article}

\def\eq#1{(\ref{#1})}
\def\be{\begin{eqnarray}}
\def\ee{\end{eqnarray}}
\def\l{\lambda}
\def\t{\tau}
\def\nn{\nonumber}
\def\Z{{\bf Z}}
\def\.{\!\cdot\!}
\def\labels#1{\label{#1}}
\def\h{{1\over 2}}
\def\r2{\sqrt{2}}

\begin{document}

\title{NEUTRINO OSCILLATIONS IN EXTRA DIMENSIONS}

\author{C.S. Lam}

\address{Department of Physics, McGill University\\ 
Montreal, QC, Canada H3A 2T8\\E-mail: Lam@physics.mcgill.ca}


\maketitle\abstracts{The characteristics and phenomenology of neutrino
oscillation in extra dimensions are briefly reviewed}

\section{Introduction}
The spacetime as we know it is four dimensional. 
A fifth dimension was postulated  by Kaluza
and Klein (KK) \cite{KK} in the 1920's,  in an attempt
to unify electromagnetism and gravity. 
If a fifth dimension exists, it must be very small or else
it would have been seen. To estimate how small that is, recall that
the allowed wave numbers in the fifth dimension 
are integral multiples of $1/R$, the inverse radius of that extra
dimension. Excited wave numbers are seen as massive
particles in our four-dimensional world, with a mass gap $1/R$. 
Since no such KK particles are known
up to about 1 TeV, the radius $R$ must be smaller than
1 (TeV)$^{-1}=2\times 10^{-19}$ m. This is the picture of the fifth
dimension until a few years ago.

Recently a different view emerged 
\cite{ADD} which allows the extra dimensions to be large.
Partly motivated by the discovery of D-branes in string theory 
\cite{POL}, it postulates all
 Standard Model (SM) particles to be permanently
confined to our four-dimensional world, the so-called `3-brane'.
Being confined they can have no KK excitations,
so the previous bound on $R$ is no longer valid. On the other hand,
SM singlets such as gravity and right-handed neutrinos
are allowed to wander into the
extra-dimensional world, `the bulk'. 
As a consequence,
the inverse-square law of gravitational force
is modified at separations less
than $R$. Such a deviation has not been detected down to submillimeter
separations \cite{EXPT}, from which we obtain
 an upper bound of $R$ to be about $10^{-4}$ m, 
not the much smaller $10^{-19}$ m discussed
before. Moreover, if there are $n$ extra dimensions, the
fundamental energy scale will change from $M_P\sim 10^{19}$ GeV
to $M_f=[M_P^2/(2\pi R)^n]^{1/(n+2)}$.
For $n=2$ and $R=0.4$ mm, this can be as low as 1 TeV, leading to many
observable consequences. However, if $n=1$, $M_f=2.3\times 10^8$ GeV
remains quite beyond our present capability to reach,
then neutrino physics is probably the only way to detect the
extra dimension. This is why neutrino
physics is so important and so interesting in this connection. In this talk 
I will  discuss some of the
consequences for neutrino oscillation when there is only one large
extra dimension present. The small extra dimensions decouple and will not
be taken into account. 
For simplicity I will assume the fifth dimension
to be flat, though a curved scenario is also interesting \cite{RS}.

\section{Neutrinos Are Different} 
Neutrino is unique as an extra-dimensional
probe because the right-handed neutrino is a
SM singlet. A large radius $R$ in the extra dimension creates a small
energy scale $1/R$ that only neutrinos can see, which may be
why neutrinos
have such exceptionally small masses. For $R=0.4$ mm, the energy
$1/R$ is half a millivolt. Whether this is really the natural energy scale
is unimportant for what follows because I will treat the neutrino masses
as parameters.

A right-handed neutrino roaming in the bulk is derived from
a 5-dimensional Dirac field, and is sterile. It
gives rise to 
a KK  tower of left-handed and right-handed
neutrinos. The minimum content of a five-dimensional theory
consists of three active brane neutrinos and a sterile bulk neutrino.
Depending on the model, there may of course be more brane neutrinos
which are sterile, and/or more bulk neutrinos.

Quark mixing via the CKM matrix is small, but at least some of
the mixings for
active neutrinos is large. One might be able to expalin this difference  if
 the additional amount of neutrino mixing is 
derived from their coupling 
to the bulk neutrino(s). 

In the presence of a bulk neutrino in a large extra dimension,
neutrino oscillations possess special
characteristics.
The infinite number of states in the KK tower generally leads to a
very complicated oscillation pattern. Besides, these neutrinos are 
sterile so they have neither charge nor neutral current interactions
with matter. These are features detectable 
from neutrino oscillation experiments.

I will discuss two simple models \cite{DDG,ADDM}
 in some detail to illustrate
these characteristics. These models are instructive though
possibly too simple to be
realistic. I will also discuss more complicated models later in the section
on phenomenology.

\section{Mixings and Oscillations}
Before discussing these models 
 let me first review the usual formalism of neutrino oscillations.
Let $\nu_e,\nu_\mu,\nu_\tau$
be the active neutrinos in the flavor basis, namely, those obtained
directly from the charged current decays of $W^+$. Through their
mutual interaction and perhaps interactions with sterile neutrinos
that may be present, they mix to form 
the mass eigenstates $\tilde\nu_\lambda$:
\be
\nu_i=\sum_\lambda U_{i\lambda}\tilde\nu_\lambda\quad (i=e,\mu,\tau).
\labels{mix}\ee
There are as many eigenvalues $\lambda$ as there are total number of
neutrinos, active and sterile ones both counted. The sum is taken over all
these eigenvalues, which consists of $3+s$ terms in a four-dimensional
world with 3 active and $s$ sterile neutrinos. 
In a five-dimensional world, there are always an infinite number of 
sterile neutrinos so it is an infinite sum.

The transition probability  into brane species $i$,
after the incoming neutrino of species $j$ and energy $E$ has 
traversed a distance $L$, is equal to ${\cal P}_{ij}(\tau)=
|{\cal A}_{ij}(\tau)|^2$, where $\tau=L/2ER^2$. 
I will use $1/R$ as the basic energy unit and keep all
the other parameters dimensionless. The transition
amplitude is given by
\be
{\cal A}_{ij}(\tau)=\sum_\lambda U^*_{i\lambda}U_{j\lambda}e^{-i
\lambda^2\tau},\labels{amp}\ee
so the transition probability is
\be
{\cal P}_{ij}(\t)&=&1-2\sum_{\l''>\l'}\Bigg[ 2\ {\rm Re}\bigg(U^*_{i\l'}U_{j\l'}U_{i\l''}U^*_{j\l''}\bigg)
\sin^2\bigg(\h\Delta\l^2\ \t\bigg)\nn\\
&&\hskip1.7cm +\ {\rm Im}\bigg(U^*_{i\l'}U_{j\l'}U_{i\l''}U^*_{j\l''}\bigg)
\sin(\Delta\l^2\ \t)\Bigg],\label{prob}\ee
where $\Delta\l^2={\l''}^2-{\l'}^2$.
It is clear from these formulas that the oscillation pattern gets more
and more complicated as the number of eigenvalues $\lambda$ increases.
In the presence of  an extra dimension,
this number is always infinite, so the
oscillation pattern is very complcated indeed. An exception occurs
when the widths of the KK resonances are large compared to their 
separations, in which case the resonances all merge into a continuum
background and the oscillation pattern becomes relatively
 simple again. This situation will be discussed in Sec.~4.4.

\section{Two Simple Models}
I will discuss two simple models, DDG \cite{DDG} and  ADDM \cite{ADDM},
slightly generalized from their original forms.
Both are assumed to contain
 one Dirac bulk neutrino, as well as $f$ 
left-handed brane neutrinos coupled to the bulk neutrino
via Dirac mass terms proportional to  $d_i/R$. These two models
 differ in that
lepton number is conserved in the ADDM model, 
but is violated by the presence of Majorana masses $m_i/R$ 
 in the DDG model. The parameters $d_i,m_i$ are assumed to
be real but otherwise completely arbitrary.
In units of $1/R$, 
the neutrino mass matrix in the ADDM model \cite{ADDM} is 
\be M'=\pmatrix{d_1&\r2 d_1&\r2 d_1&\r2 d_1&\r2 d_1&\cdots\cr
                d_2&\r2 d_2&\r2 d_2&\r2 d_2&\r2 d_2&\cdots\cr
                \cdots&&\cdots&&\cdots&\cdots\cr
                d_f&\r2 d_f&\r2 d_f&\r2 d_f&\r2 d_f&\cdots\cr
                   0&1&0&0&0&\cdots\cr
                   0&0&2&0&0&\cdots\cr
                    0&0&0&3&0&\cdots\cr
                    0&0&0&0&4&\cdots\cr
               \cdots&&\cdots&&\cdots&\cdots\cr}.\labels{addm}\ee
The rows and columns of this matrix are labelled respectively by the left-handed 
and the right-handed neutrinos. The brane neutrinos are purely left-handed, but the KK
tower of bulk neutrino contains both left-handed and right-handed components.
They have masses $n/R$, with $n=0,1,2,\cdots$. The $n=0$ mode is special,
because its left-handed component is decoupled from everything else, so it does
not appear in the mass matrix. Its right-handed component does couple
to the $f$ brane neutrinos and it occupies the first column of the matrix.
In short, the columns are labelled by the $n\ge 0$ modes but the rows
are labelled by the $f$ brane neutrinos, followed by the $n>0$ modes. 

The brane neutrinos of the DDG model \cite{DDG}  are Majorana
so rows and columns of its mass matrix are both labelled 
by all the brane neutrinos and all the modes of the bulk neutrinos.
In units of $1/R$, the mass matrix of this model is
\be M=\pmatrix{m_1&0&\cdots&0&d_1&d_1&d_1&d_1&d_1&\cdots\cr
               0&m_2&\cdots&0&d_2&d_2&d_2&d_2&d_2&\cdots\cr
               \cdots&&\cdots&&&&&&&\cdots\cr
                 0&0&\cdots&m_f&d_f&d_f&d_f&d_f&d_f&\cdots\cr
               d_1&d_2&\cdots&d_f&0 &0 &0 &0 &0 &\cdots\cr
               d_1&d_2&\cdots&d_f&0 &1 &0 &0 &0 &\cdots\cr
               d_1&d_2&\cdots&d_f&0 &0 &-1 &0 &0 &\cdots\cr
               d_1&d_2&\cdots&d_f&0 &0 &0 &2 &0 &\cdots\cr
               d_1&d_2&\cdots&d_f&0 &0 &0 &0 &-2 &\cdots\cr
               \cdots&&\cdots&&&&&&&\cdots\cr}.\labels{mm}\ee
Note that the $n$th KK mode of the ADDM model is a linear
 combination of the $\pm n$ modes of the DDG model. Only one combination
occurs in the mass matrix $M'$ because the other combination decouples.

We shall assume the $2f$ parameters $m_i$ and $d_i$ to be real, in which
case the mass matrix $M$ in the DDG model 
is real and symmetrical. The unitary mixing matrix $U$ (actually
real orthogonal in this case),
and the eigenmasses $\l/R$, can both be obtained by diagonalizing $M$.
The resulting eigenvalues $\lambda$
satisfy the characteristic equation
\be
{1\over\pi}\tan(\pi\l)=
\sum_{i=1}^f{d_i^2\over\l-m_i},\labels{charac}\ee
 whose graphical solution 
is illustrated in Fig.~1 for $f=3$ and two different values of $d^2
\equiv \sum_{i=1}^fd_i^2$. 
We shall denote $\tan(\pi\l)/\pi$ by $\eta(\l)$, and the sum on
the right-hand side of \eq{charac} by $d^2r(\l)$.
We shall also write $d_i=de_i$.
Therefore $\sum_ie_i^2=1$ and $r(\l)=\sum_ie_i^2/(\l-m_i)$.

The components of the corresponding eigenvector  
\be 
\ell=(w_1,w_2,\cdots,w_f,v_0,v_1,v_{-1},v_2,v_{-2},\cdots)^T\ee
are
\be
w_i(\l)&=&d_i/(\l-m_i)\quad(i=1,2,\cdots,f),\nn\\
v_n(\l)&=&\eta(\l)/(\l-n)\quad (n\in {\bf Z}).\labels{vec}\ee
The norm of the  eigenvector is 
\be
B^2(\l)=\ell^T\.\ell&=&\sum_id_i^2/(\l-m_i)^2+1/\cos^2(\pi\l)\nn\\
&=&d^2s(\l)+[1+d^4\pi^2r^2(\l)],\labels{norm}\ee
where \eq{charac} has been used. 
The first term comes from $\sum_iw_i^2$, also denoted
as $d^2s(\l)$, and the second term comes from $\sum_nv_n^2$. The 
unitary mixing
matrix $U_{i\l}$ in eqs.~\eq{mix} and \eq{amp} is given by 
\be
U_{i\l}=w_i(\l)/B(\l).\labels{u}\ee

For the ADDM model, whose mass matrix $M'$ is not symmetrical,
we need to compute the eigenvalues $\l^2$
and the eigenvectors $\ell=(w_1,\cdots,w_f,v_1,v_2,\cdots)^T$
of $M'(M')^T$. The characteristic equation and the eigenvector
components are once again given
by \eq{charac} and \eq{vec}, if we set $m_i=0$ and $n\ge 1$.
In addition, there are also $f-1$ eigenvectors with $\l=0, v_n=0,$
and $(w_1,\cdots, w_f)$ orthogonal to $(d_1,\cdots,d_f)$.

\subsection{Oscillation pattern}
We are now in a position to examine the consequence of this eigenstructure on the survival probablility ${\cal P}_{ii}(\t)$ given in \eq{prob}.
For the DDG model we will assume $m_i$ to be neither integers nor half
integers, and all distinct.
Since phases are absent in these models, only the real part contributes,
hence
\be
{\cal P}_{ii}(\t)&=&1-
\sum_{\l''>\l'}(U_{i\l'}U_{i\l''})^2
\sin^2(\Delta\l^2\ \t/2).\labels{surv}\ee

\subsection{Weak-coupling limit}

When $d_i=0$, the matrix \eq{mm} is diagonal, with brane eigenvalues
$m_i$ and bulk eigenvalues $n\in\Z$. 
For couplings $d$ much smaller than the separation
between any of these free eigenvalues, the eigenvalues are shifted 
very little so it is convenient to label them
by the unshifted ones: $\l_i\sim m_i$ and $\l_n\sim n$. The eigenvectors
and their norms can be computed using perturbation theory from \eq{charac}
to \eq{u}, to yield
\be
U_{i\l_i}&\simeq&1\nn\\
U_{i\l_j}&\simeq& d^2e_ie_j/\eta(m_j)(m_j-m_i)\nn\\
U_{i\l_n}&\simeq& d^2r(\l_n)/(n-m_i).\labels{uweak}\ee 
Assuming $r(\l_n)$ and $\eta(m_j)$ to be of order unity, 
it folows that the small 
elements $U_{i\l_j}$ and $U_{i\l_n}$ are of order $d^2$,
and inversely proportional to the distance between the
unshifted eigenvalues and $m_i$. 

Appling this result to \eq{surv}, we can approximate it by
\be
{\cal P}_{ij}(\t)&\simeq&1-\sum_{\l}U_{i\l}^2\sin^2(\Delta\l^2\ \t/2).
\labels{surv2}\ee
Thus the high `frequency' components with large
$\Delta\l^2\simeq \l^2-m_i^2$ are weakened by the factor
$1/(\l-m_i)^2$.

These results, derived for the DDG model, is also valid for the ADDM
model with minor changes.  
Since $f-1$ of the brane neutrinos decoupled from the rest of
the ADDM model and can be taken into account easily, 
let us just discuss the case $f=1$. In the absence of coupling,
the eigenvalue is $\l_b=0$ for the brane and $\l_n=n=1,2,\cdots$
for the bulk. For weak coupling, $\l_b\simeq d$, $U_{1\l_b}
\simeq 1$, and $U_{1\l_n}\simeq d^2/n^2$.

\subsection{Intermediate coupling}
In this case many KK modes are excited. The oscillation pattern 
gets very wiggly and very 
complicated, as illustrated in Fig.~2 from a numerical 
calculation taken from Ref.~[8].
It seems unlikely that this is the situation phenomenologically
unless we are so unlucky to have missed all these `irregular' 
patterns so far.

\subsection{Strong coupling}
One might think that if intermediate coupling is difficult to analyse,
strong coupling will be impossible. This turns out not to be the case
because the widths of the eigenstates become so large
that they overlap with one another to eliminate most of the wiggly
and complicated structure seen in intermediate couplings. 
 In fact, perhaps somewhat surprisingly, the problem
becomes  exactly soluble \cite{LN} even for the DDG model.

To explain how that comes about let us
first examine the eigenvalues in the strong coupling limit
($d\gg m_i,1$). For any $d>0$, we see from
Fig.~1 that $r(\l)$ is a decreasing function of $\l$, with
poles at $m_i$, but $\tan(\pi\l)$ is an increasing
function of $\l$, with poles at half integers. Consequently, there is
one and only one solution of \eq{charac} in any interval bounded by
a neighboring pair of poles.  
As $d$ is increased, the solution 
slides towards bigger magnitude of $\tan(\pi\l)$, {\it i.e.}, 
$\l$ increases if $r(\l)>0$ and decreases if $r(\l)<0$.
When $d\to\infty$, an eigenvalue must end up at the boundary of 
the interval, or at a zero of $r(\l)$. It is not difficult to show
that the eigenvalues then consist of all half integers, plus the $f-1$ 
zeros of $r(\l)$.
We denote the former by $\l'_n=n+\h\ (n\in\Z)$, and the
latter by $\l_\ell\ (2\le\ell\le f)$. $\l'_n$
will be referred to as the {\it regular}
 eigenvalues, and $\l_{\ell}$ will be referred to 
as the {\it isolated} eigenvalues. Note that for large $d$, the brane
and bulk neutrinos are quite thoroughly mixed up so we can no longer
tell whether an eigenvector is more brane-like or more bulk-like. 
In particular,
if we trace an isolated eigenvalue continuously back to $d=0$,
depending on the precise values of $e_i$ and $m_i$, sometimes it ends up as
a brane eigenvalue $m_i$, and sometimes it ends up as a bulk 
eigenvalue $n$.

Next let us look at $B^2(\l)$ and $U_{i\l}$. They behave differently
 for the two kinds of  eigenvalues, because
$\tan(\pi\l)$ is infinite for the regular eigenvalues
 and finite for the isolated eigenvalues.
For this reason,
it is more convenient to write \eq{norm} for the regular
eigenvalues in the form $B^2(\l)=
d^2s(\l)+1+d^2\tan^2(\pi\l)$, from which we see that $B^2(\l)=O(d^2)$
and $U_{j\l}$ is finite in the large-$d$ limit.
For regular eigenvalues, $r(\l)$ is non-zero, so we see from
\eq{norm} that $B(\l)$ is of order $d^4$. This means that $U_{j\l}$
vanishes in the limit for every $\l$ of order unity. However, it does not mean
that the sum in \eq{amp} over all regular eigenvalues is zero, because there is an
infinite number of regular eigenvalues. In fact, for large $\l$ of order
of $d^2$, we have $r(\l)\simeq 1/\l$ and $s(\l)\simeq 1/\l^2$, so 
$B^2(\l)\simeq 1+d^2(1+\pi^2d^2)/\l^2\equiv 1+K^2/\l^2$ is finite for
fixed $d^2/\l$ in the strong coupling limit. 
As a result, the sum in \eq{amp} over the regular
eigenvalues can be replaced by the sum between $n=\pm N$ and
$\pm\infty$, for any finite number $N$. We will choose
$N$ so that $\l'_N\gg 1,m_i$, then
this sum can in turn be replaced by an integral, yielding
\be
{\cal A}^{reg}_{ij}&=&\sum_{n=-\infty}^\infty
U^*_{i\l_n}U_{j\l_n}r^{-i\l_n^2\t}\nn\\ 
&=&e_ie_j\int_{-\infty}^\infty d\l{d^2e^{-i\l^2\t}\over\l^2+K^2}\nn\\
&=&e_ie_jg(K^2\t),\labels{areg}\ee
where
\be
g(x)&\equiv&{1\over\pi}\int_{-\infty}^\infty du{e^{-iu^2x}\over u^2+1}
\labels{g}\ee
begins at $g(0)=1$. Its absolute value decreases monotonically in
$x$, approaching $g(x)\simeq (1-i)/\sqrt{2\pi x}$ for large $x$.

This leaves the sum over isolated eigenvalues. Since $U_{i\l}$ is finite
for these eigenvalues, the transition in the strong coupling limit is
amplitude 
\be{\cal A}_{ij}(\t)&=&e_ie_j\Bigg[g(K^2\t)
+\sum_{\ell=2}^f {e^{-i\l_\ell^2\tau}\over s(\l_\ell)(\l_\ell-m_i)
(\l_\ell-m_j)}\Bigg]
\nn\\
&\equiv&V^*_{i1}V_{j1}g(K^2\t)+\sum_{\ell=2}^fV^*_{i\ell}V_{j\ell}
e^{-i\l_\ell^2\tau}
\labels{ampfin}\ee
It can be verified that the $f\times f$ matrix $V$ is unitary 
(actually real orthogonal), so that ${\cal A}_{ij}(0)=\delta_{ij}$
as it should. If the function $g(K^2\t)$ were replaced by $e^{-i\l_1^2\t}$
(for some $\l_1$), then \eq{ampfin} would just be
 the usual transition amplitude
of $f$ fermions, with a mixing matrix $V_{ij}$ and no bulk
neutrino. In that case unitarity of $V$ guarantees the 
conservation of (brane) neutrino probability, namely, ${\cal P}_j(\t)\equiv
\sum_{i=1}^f{\cal P}_{ij}(\t)=1$ for all $\t$. With the presence of $g(K^2\t)$
in \eq{ampfin}, there is a leakage into the bulk so the total brane
neutrino probability decreases with `time', as in
\be
{\cal P}_j(\t)=1-|e_j|^2\{1-|g(K^2\t)|^2\}.\labels{totprob}\ee
This reaches an asymptotic amount $1-|e_j|^2$ for large $\t$. Since 
$\sum_j|e_j|^2=1$, the average leakage per brane channel is $1/f$.

\section{Phenomenology}
According to the Super-Kamiokande analysis of their data \cite{SK},
the mixing angle for atmospheric neutrino is large, and the mixing with
sterile neutrino is small. Their data on solar
neutrino favor
the large angle MSW solution LMA, without much mixing with a
sterile neutrino. However, this conclusion
on solar neutrino depends somewhat how the data are analysed.
The small angle (SMA) and the vacuum oscillation (VAC) solutions, with or
without
a large amount of sterile neutrino present, are not completely ruled out
\cite{BGP}, even with the recent charge current data
from SNO \cite{SNO} taken into account.

There have been several attempts to understand the data from
bulk neutrino models \cite{CLGFMV,WK}.
Unfortunately the data pose considerable difficulties on the 
simple DDG/ADDM model, which contains
only one bulk neutrino and has no direct mixing between the active neutrinos.
In the weak coupling limit when the mixing is small, 
there is no way this simple model
can be applied to the atmospheric neutrino problem. 
Motivated by the 2+2 solution
introduced to explain also the LSND data \cite{LSND}, 
in which $(\nu_\mu,\nu_\t)$ and
$(\nu_e,\nu_s)$ form nearly degenerate doublets separated by about 1 eV, 
and the solar neutrino oscillates mostly to the sterile neutrino,
attempts have been made to use the DDG/AGGM model  to 
explain the solar neutrino
problem. However, the 2+2 solution contains one sterile
neutrino $\nu_s$, but the DDG/ADDM model contains an infinite tower of them.
With more sterile neutrinos present the oscillation pattern becomes more
complicated, and this can be utilized to cure the wrong energy dependence
in the Super-K data caused by a single $\nu_s$. Two groups have succeeded in
fitting the solar neutrino data, before the SNO data became available
anyway,
either with a $f=1$ DDG model \cite{MOH}, or by introducing a mass for the
bulk neutrino in the $f=1$ ADDM model \cite{LRRR}. To be able to explain also
the atmospheric and perhaps the LSND data, one must modify the 
DDG/ADDM model, by introducing direct mixing between the active neutrinos,
by introducing three bulk neutrinos, or by some other means.

One might think that the large mixing of atmospheric neutrino, and perhaps
also of the solar neutrino, could possibly be explained
 by the DDG/ADDM model in the presence of a
strong coupling. However, strong coupling also diverts a considerable
amount of active neutrino
flux into sterile neutrinos (the first term within the square brackets
in \eq{ampfin}), which is forbidden by the data. Nevertheless, 
if the diversion
comes mostly from an incoming $\nu_\t$ beam, for which there is no data,
it is conceivable that it might still work. In other words, we might
attempt to make $e_\t$ in \eq{totprob} close to 1 and $e_e, e_\mu$ small,
in which case there is not much leakage of the solar and atmospheric
neutrino fluxes into the sterile channels. Unfortunately this does not
work for the $f=3$ DDG model, because with the vanishing of the first
term of \eq{ampfin} for $\t\gg 1/K^2$, there remains only $f-1=2$ terms
in \eq{ampfin} and only one square-mass difference, impossible to explain
both the atomospheric and the solar oscillations. Whether there is a way
to make this work for $f\ge 4$ has not yet been analysed.

\newpage
\begin{figure}[ht]
\includegraphics{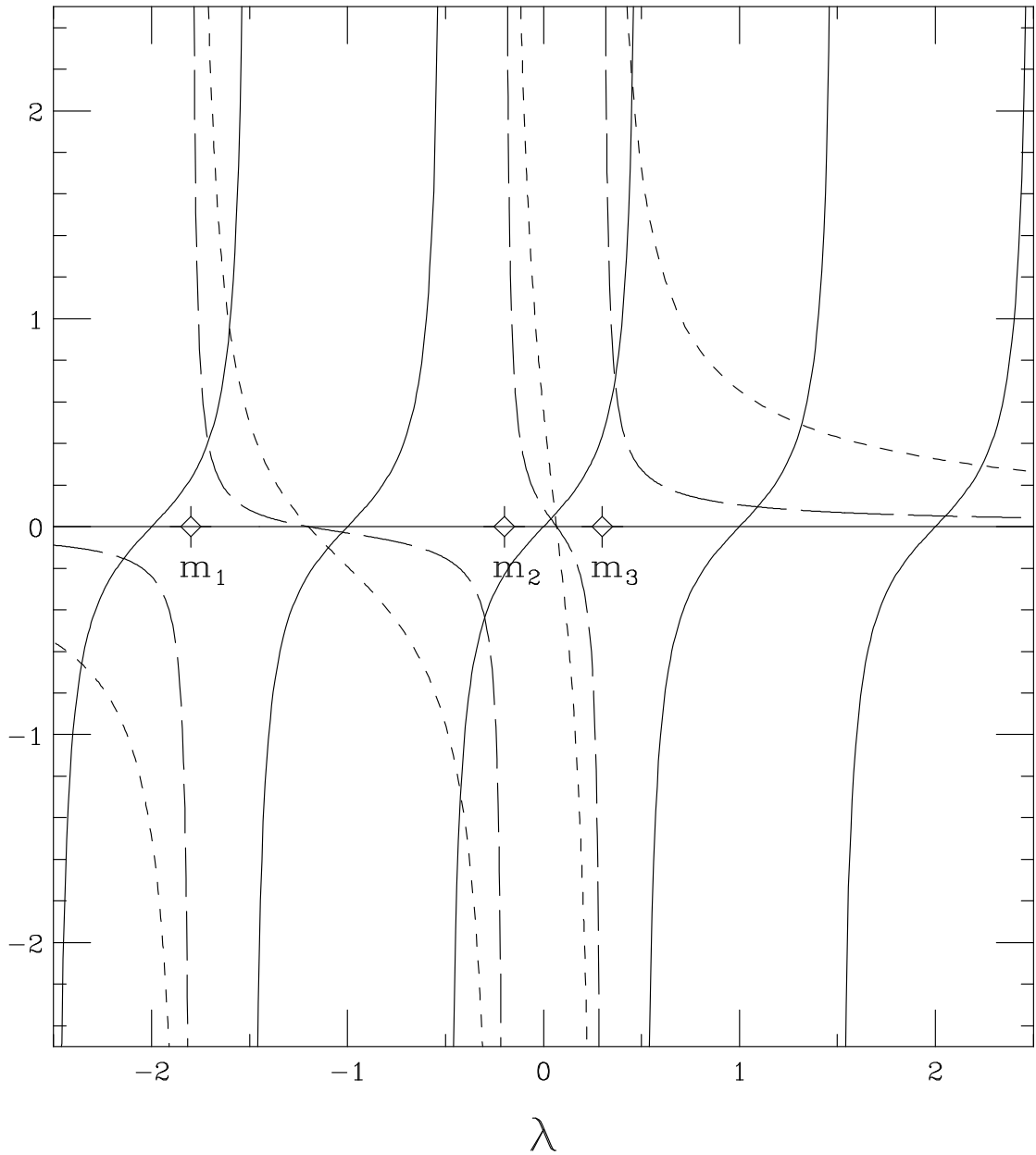}
\vspace{14cm}
\caption{Graphical solutions of the characteristic equation \eq{charac}.
The left hand side of \eq{charac} is represented by the solid curve; the
right hand side is represented by the dashed curves, with $(m_1,m_2,m_3)=
(-1.8,-0.2,0.3)$. The one with the short dash has the larger coupling
constant $d$ than the one with the long dash. Note that the solution
$\l$ moves up and right along the solid curve for increasing $d$
when the value on both
sides of \eq{charac} is positive, and moves down and left along the solid
curve when the value is negative.}
\end{figure}

\newpage
\begin{figure}[ht]
\includegraphics{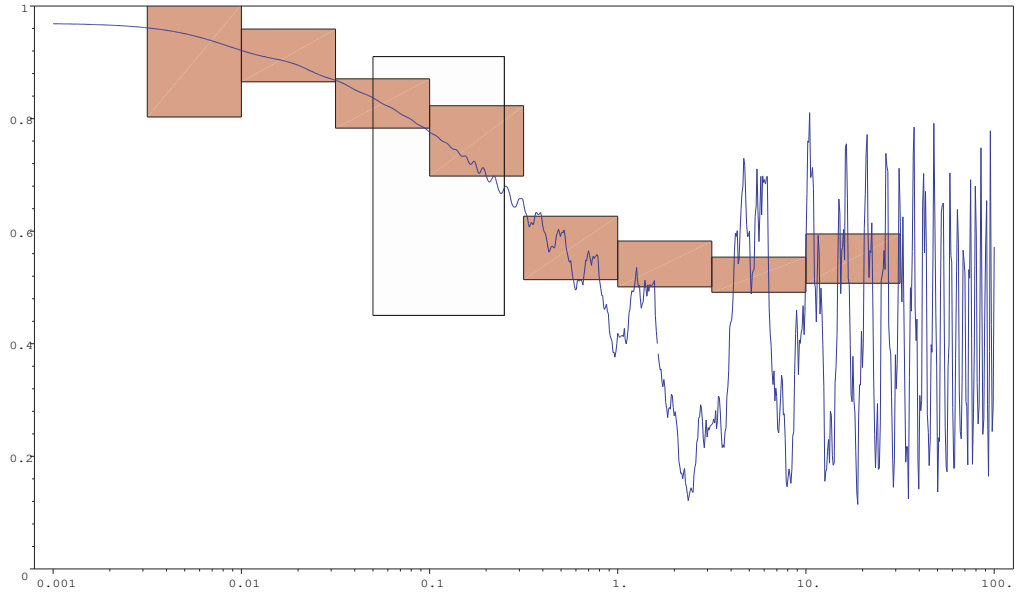}
\vspace{8cm}
\caption{Taken from Fig.~5 of 
Ref.~[8] to show the wiggly
pattern in the $L/E$ dependence,
with a moderate coupling to the infinite tower of KK neutrinos.
The boxes are the Super-Kamiokande and K2K results.}
\end{figure}

\end{document}